\documentclass{PoS}
\usepackage{amsmath,amssymb}
\usepackage{dsfont}
\usepackage{slashed}
\usepackage{verbatim}
\usepackage{wrapfig}
\usepackage{mciteplus}

\renewcommand{\d}{\textmd{d}}
\newcommand{\be}{\begin{equation}}
\newcommand{\ee}{\end{equation}}

\newcommand{\Z}{Z} 
\newcommand{\F}{F} 
\newcommand{\PP}{\mathcal{P}}

\def\dena{A(\mathcal{E})}
\def\dbna{A(\mathcal{B})}
\def\df{\mathcal{C}}

\newcommand{\fracp}[2]{\frac{\partial #1}{\partial #2}}

\newcommand{\Regensburg}{Institute for Theoretical Physics,
Universit\"at Regensburg, D-93040 Regensburg, Germany.}

\newcommand{\Mumbai}{Dept. of Theoretical Physics, Tata Institute of Fundamental Research, Homi Bhabha Road, Mumbai 400005, India.}

\title{Magnetization and pressures at nonzero magnetic fields in QCD}

\ShortTitle{Magnetization and pressures at nonzero magnetic fields in QCD}

\author{
G.~S.~Bali$^{1,2}$, F.~Bruckmann$^{1}$, \speaker{G.~Endr\H{o}di}$^{1}$,
A.~Sch\"afer$^1$ \\
        E-mail: \email{gergely.endrodi@physik.uni-regensburg.de}
}

\footnotetext[1]{\Regensburg}
\footnotetext[2]{\Mumbai}

\abstract{
Two approaches are pursued to determine the magnetization of the QCD vacuum at zero 
and nonzero temperatures using lattice simulations. 
The first 
method builds on pressure anisotropies which are induced by the magnetic field on the lattice. 
The second approach is a novel, generalized version of the integral method, which exploits the 
independence of the theory on $B$ for asymptotically large quark masses. 
Both approaches give consistent results and confirm that the QCD medium is paramagnetic. 
Finally, an interesting relation between QCD paramagnetism, magnetic catalysis of the QCD condensate 
and the QED $\beta$-function is pointed out.
}

\FullConference{31st International Symposium on Lattice Field Theory LATTICE 2013\\
		 July 29 -- August 3, 2013\\
		 Mainz, Germany}

\begin{document}

\section{Introduction}

Strong (electro)magnetic fields are very efficient probes of the thermal vacuum of 
Quantum Chromodynamics (QCD). Given the 
electrically charged nature of quarks and the electric neutrality of gluons, 
magnetic fields effectively disentangle the elementary particles of the theory, 
if the field competes in strength with the QCD interactions.
This implies that a magnetic field of the order of several $m_\pi^2$ is necessary. 
Such a strong magnetic field can indeed induce several new phenomena, for example it 
generates Lorentz-symmetry-breaking expectation values, it affects 
chiral symmetry breaking and restoration and (de)confinement, and it considerably changes 
the hadron spectrum. For recent reviews on the subject, see, 
e.g., Refs.~\cite{Kharzeev:2012ph,Bali:2013cf}.  
Remarkably, magnetic fields of this strength occur in nature (in magnetars~\cite{Duncan:1992hi} 
and during the electroweak 
epoch of the evolution of the early universe~\cite{Grasso:2000wj}) and in experiments 
(relativistic heavy-ion collisions~\cite{Skokov:2009qp}), 
making the study of QCD with magnetic fields relevant for these systems.

All information about the magnetic properties of QCD is contained in the free energy of 
the system, which is given in terms of the partition function as $\F=-T\cdot \log\Z$. 
The response to the magnetic field is given by the magnetization density,
\be
M=-\frac{1}{V}\frac{\partial \F }{\partial (eB)},
\label{eq:Mdef}
\ee
where 
the magnetic field is 
given in units of the elementary charge $e>0$. 
In particular, a positive magnetization implies that the QCD vacuum as a medium is a paramagnet, 
whereas the opposite sign corresponds to diamagnetism. 
While paramagnets decrease their free energy when exposed to an external field, for diamagnets 
it is energetically favorable to repel the magnetic field. 
Thus, the sign of $M$ is clearly a fundamental characteristic of the thermal QCD vacuum. 

\begin{figure}[t]
 \centering
 \vspace*{-.3cm}
 \includegraphics[width=3.5cm]{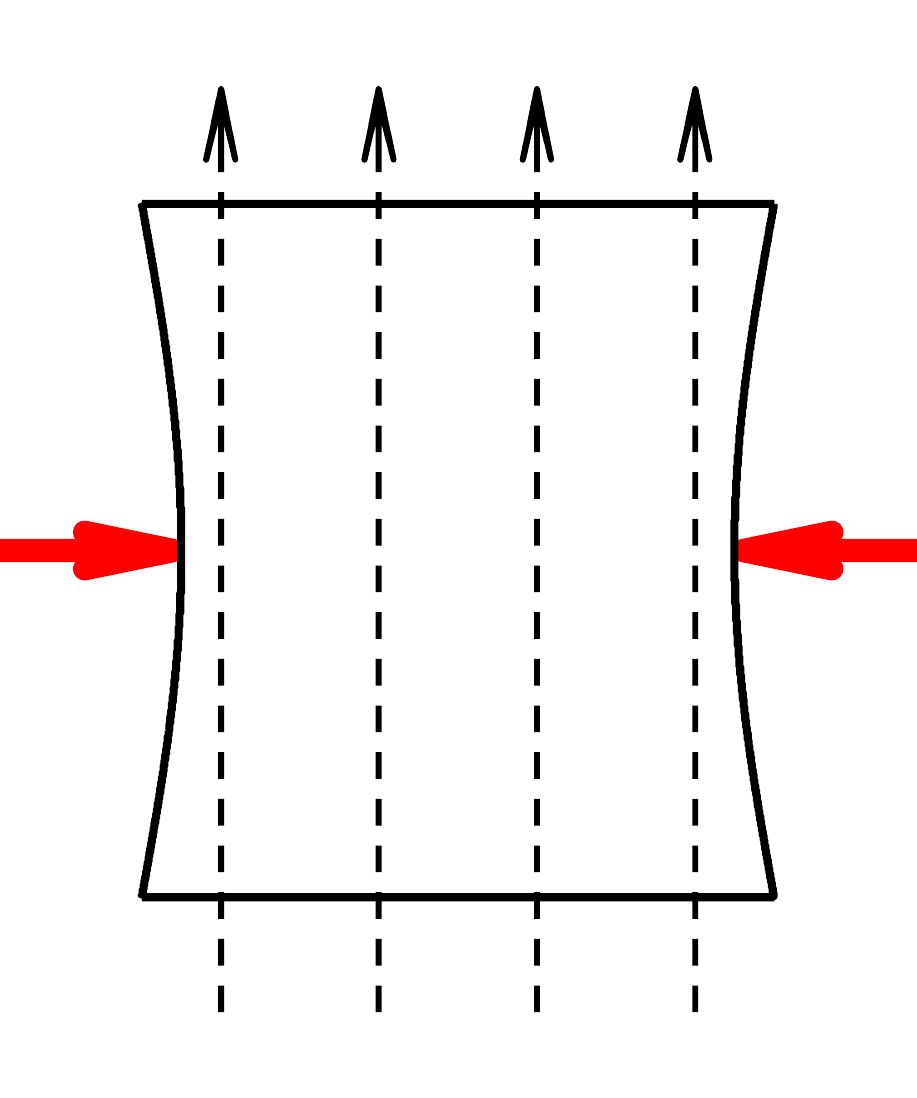} \quad\quad\quad\quad
 \includegraphics[width=3.5cm]{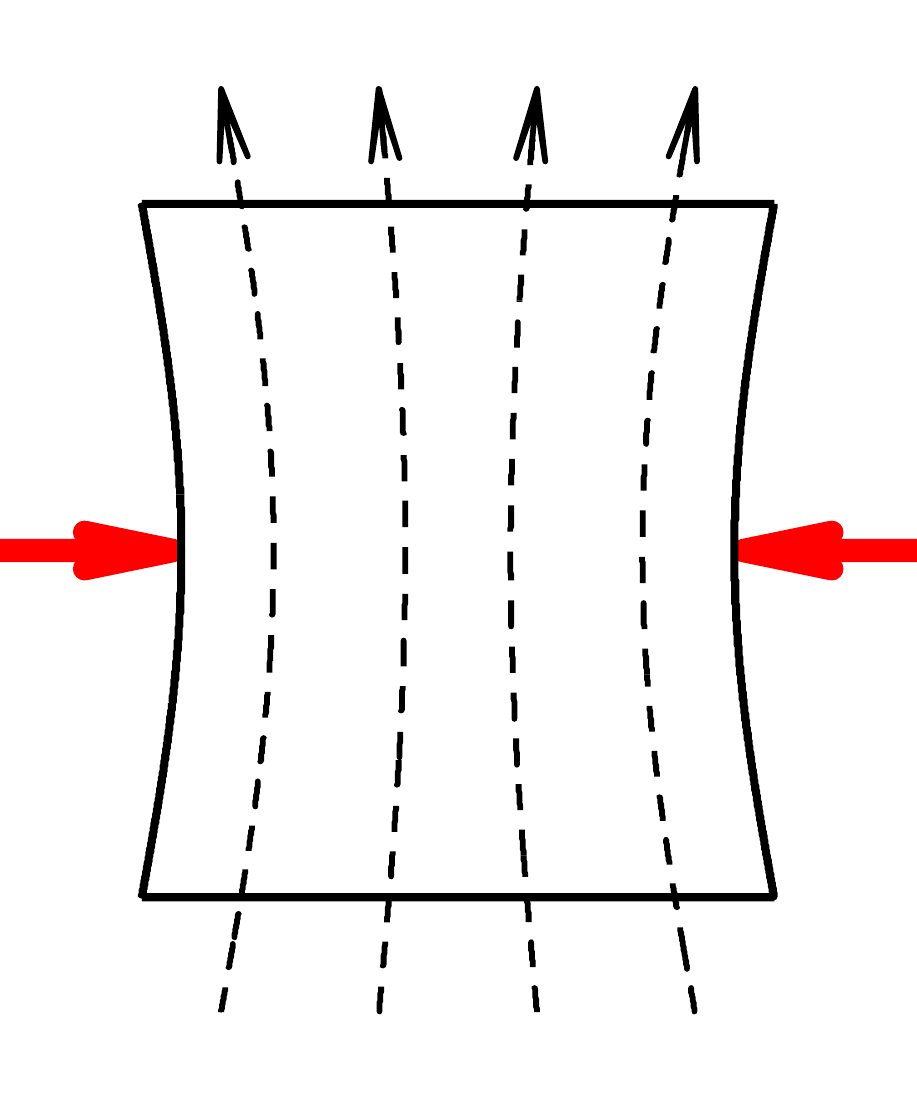}
 \vspace*{-.3cm}
 \caption{ \label{fig:illustr}
  Two possible scenarios for defining the transverse pressures in the presence of a magnetic field: 
keeping either the field constant ($B$-scheme, left panel) or the flux constant 
($\Phi$-scheme, right panel).
 }
\end{figure}

The dependence of $\F$ on $B$ is also important for the determination 
of the QCD equation of state (EoS) in the presence of the magnetic field. 
For the EoS, the primary observable is the pressure. For nonzero magnetic fields, the direction 
parallel to $B$ is distinguished and, in principle, the spatial components of the pressure may 
become different. To be more specific, let us consider a finite volume of size $V=L_xL_yL_z$, 
for which the pressure components are written as
\be
p_i = -\frac{1}{V} L_i\frac{\partial \F}{\partial L_i}.
\label{eq:pidef}
\ee
Applying the magnetic field in the $z$ direction induces a magnetic flux 
$\Phi=eB\cdot L_xL_y$. To define the pressures we have to specify the trajectory 
in parameter 
space, along which the partial derivative in Eq.~(\ref{eq:pidef}) is evaluated. Two possibilities 
are, for example, to keep the magnetic field constant ($B$-scheme) or, to keep the magnetic 
flux constant ($\Phi$-scheme)~\cite{Bali:2013esa}. In the $B$-scheme, compressing the system in the transverse ($x$ 
or $y$) directions decreases the flux, whereas in the $\Phi$ scheme, the same compression 
increases the magnetic field. The two scenarios are illustrated in Fig.~\ref{fig:illustr}.

Let us consider a large homogeneous system, where the free energy is extensive, $\F(L_i,eB,T) = V\cdot 
\Omega(eB,T)$ (such an extensive free energy describes, for example, free charged particles, see, e.g., 
Ref.~\cite{Endrodi:2013cs}). 
In the $B$-scheme, the spatial extensions only appear in the prefactor, therefore, the pressures will be 
{\it isotropic}. In the $\Phi$-scheme, on the other hand, the transverse derivatives 
of $\F(L_i,\Phi,T) = V\cdot \Omega(\Phi/(L_xL_y),T)$ become different, 
resulting in {\it anisotropic} pressures. 
Since the additional dependence on $L_x$ and $L_y$ is through the $B$-dependence of $\F$, the 
anisotropy is proportional to the magnetization, Eq.~(\ref{eq:Mdef}). 
On the contrary, the longitudinal pressure $p_z$ is the same in both schemes. 
Altogether, we obtain in the two schemes
\be
p_{x,y}^{(B)} = p_z,\quad\quad\quad p_{x,y}^{(\Phi)} = p_z - M\cdot eB,
\label{eq:paniso}
\ee
and in the thermodynamic limit, the longitudinal pressure is given as $p_z=-\F/V=T/V\cdot \log\Z$. 
Let us stress that there is no `correct' scheme; which trajectory $B(L_i)$ one should choose to evaluate 
the pressures depends on the physical situation in question. The above introduced 
$B$- and $\Phi$-schemes are merely two possibilities. 
One example for the $\Phi$-scheme is a perfectly conducting plasma, where magnetic field lines are frozen in and, therefore,	 flux is conserved.

Another important aspect of the magnetic field-dependence of the free energy is its 
renormalization. Besides the $B=0$ divergences of $\F$, the magnetic field 
induces an additional divergent term of the form 
$\beta_1 \cdot (eB)^2 \cdot \log \Lambda$, where the regulator $\Lambda$ of the theory is
to be taken to infinity. The coefficient $\beta_1$ equals the leading order coefficient 
of the QED $\beta$-function (with QCD corrections). 
This is no coincidence, as this divergence stems from the electromagnetic interaction of the quarks 
with the magnetic field and can be canceled by the simultaneous renormalization of 
the electric charge $e$ and of the magnetic field $B$. For $\F$, this corresponds to 
a redefinition of the energy $B^2/2$ of the magnetic field,
\be
\frac{B^2}{2} + \beta_1 \cdot (eB)^2 \cdot \log \Lambda = \frac{B^{r2}}{2},
\label{eq:B2half}
\ee
and coincides with the usual wave-function renormalization $B^2=Z_e B^{r2}$, 
where the superscript indicates that $B^r$ is a renormalized quantity. Note that the electric 
charge renormalizes as $e^2=Z_e^{-1}e^{r2}$ and, thus, their product is invariant, $eB=e^rB^r$. 

In the on-shell scheme, the renormalization of Eq.~(\ref{eq:B2half}) 
amounts to a complete subtraction of the $\mathcal{O}((eB)^2)$ term in $\F$ at $T=0$ 
(and a similar prescription for $M$ and for the 
pressures)~\cite{Elmfors:1993wj,*Dunne:2004nc,Endrodi:2013cs},
\be
\F^r = (1-\PP)[\F], \quad\quad\quad M^r\cdot eB = (1-\PP) [M\cdot eB],
\quad\quad\quad p_z^r = (1-\PP) [p_z],
\label{eq:Frenorm}
\ee
where $\PP$ is the operator that projects out the $\mathcal{O}((eB)^2)$ term 
from a quantity $X$,
\be
\PP [X] = (eB)^2 \lim_{eB\to0} \frac{X}{(eB)^2}.
\label{eq:PPdef}
\ee
Therefore, at $T=0$ the expansion of $F^r$ in $eB$ starts with 
a quartic term. On the other hand, thermal contributions at $T>0$ induce a nonzero quadratic term as well. Note that other observables, like the condensate (see definition~(\ref{eq:densities}) below) are not affected by this renormalization.

\section{Magnetization on the lattice -- pressure anisotropies}
\label{sec:aniso}

In the rest of this talk, two independent approaches will be discussed, which can be used to 
determine the QCD magnetization in lattice simulations. 
Our numerical results were obtained using a Symanzik improved gauge action and 
$2+1$ flavors of stout smeared staggered quarks with physical masses (the setup is detailed in 
Refs.~\cite{Aoki:2005vt,Bali:2011qj}). 
On a finite lattice of size $N_s^3\times N_t$ and spacing $a$, 
the magnetic field is not arbitrary but quantized 
due to periodic boundary conditions,
\be
\Phi = eB \cdot (N_sa)^2 = 6\pi N_b, \quad\quad\quad N_b\in\mathds{Z}, \quad\quad 0\le N_b<N_s^2.
\ee
Here we have taken into account that the magnitude of the smallest quark electric charge is $e/3$. 

Flux quantization implies that the magnetization of Eq.~(\ref{eq:Mdef}) is ill-defined. 
The first approach to circumvent this problem was developed in Ref.~\cite{Bali:2013esa}. 
It is based on the fact that the lattice setup described above with constant $N_b$ realizes the $\Phi$-scheme. Thus, 
the second relation in Eq.~(\ref{eq:paniso}) can be employed to determine $M$. 
Moreover, in Ref.~\cite{Bali:2013esa} we have also shown how to write the difference between the 
pressure components as anisotropies of the lattice action, giving the relation
\be
-M\cdot eB = -(\zeta_g+\hat\zeta_g) \left[ \dbna - \dena\right] - \zeta_f \sum_f A(\df_f).
\label{eq:aniso}
\ee 
Here, $\dena$ and $\dbna$ are anisotropies of the chromoelectric and chromomagnetic parts of the 
gluonic action and $A(\df_f)$ is the anisotropy of the fermionic action for the flavor $f$ ($f=u,d,s$). 
The coefficients $\zeta_g$, $\hat\zeta_g$ and $\zeta_f$ are anisotropy renormalization coefficients, which
in principle should be determined non-perturbatively, e.g., through 
simulations on anisotropic lattices. This is a highly complicated and expensive task. 
For the moment, we substitute these coefficients by their leading perturbative values, $\zeta_g=\hat{\zeta}_g=\zeta_f=1$. 
Moreover, since we found that the right hand side of Eq.~(\ref{eq:aniso}) is dominated by the 
fermionic anisotropy, in the following we estimate the magnetization by considering 
simply $\sum_f A(\df_f)$. This amounts to a systematic error of about $10-20\%$. 
For the derivation of Eq.~(\ref{eq:aniso}) and additional details, see 
Ref.~\cite{Bali:2013esa}. 

\begin{figure}[ht!]
 \centering
 \includegraphics[width=7cm]{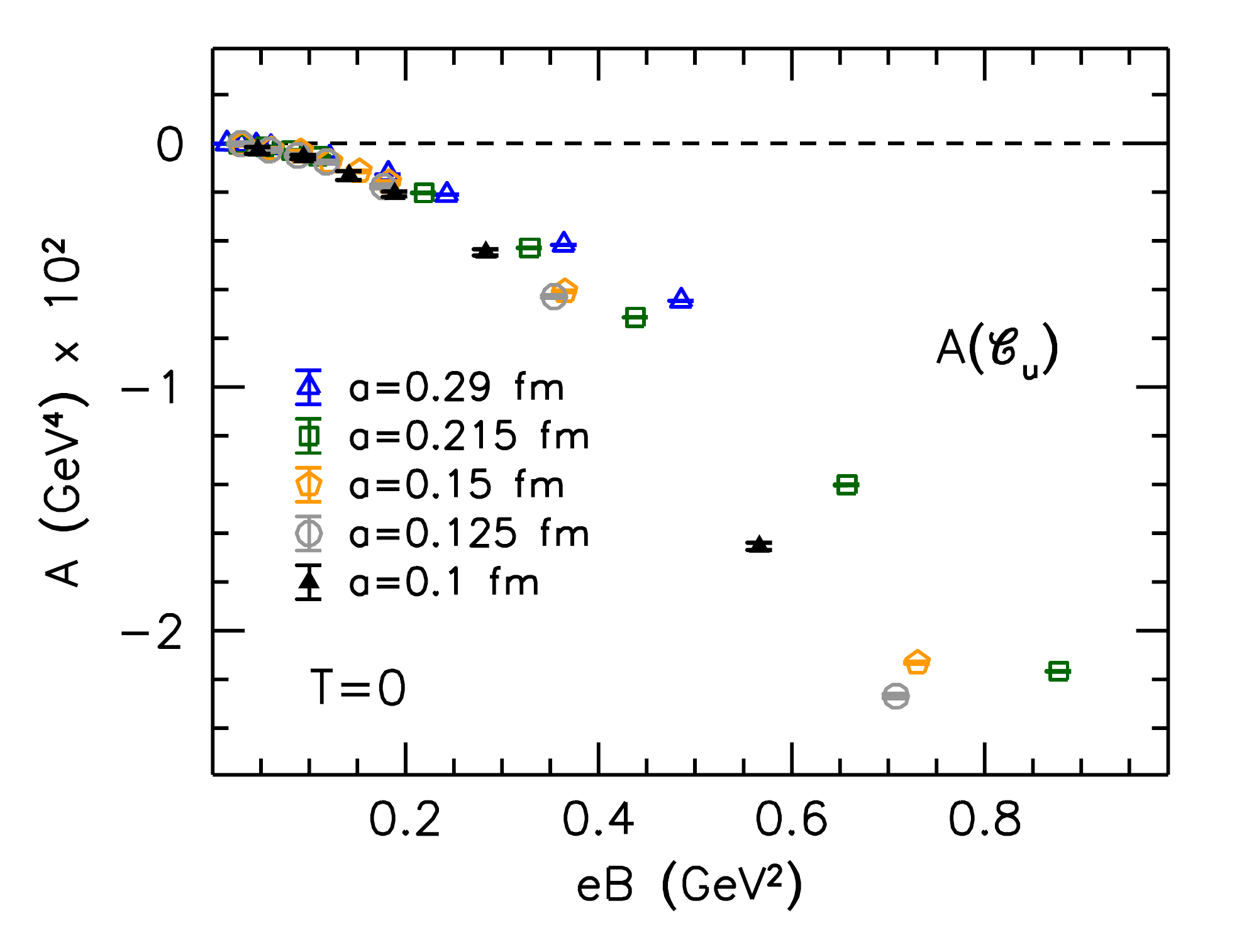} \quad\quad
 \includegraphics[width=7cm]{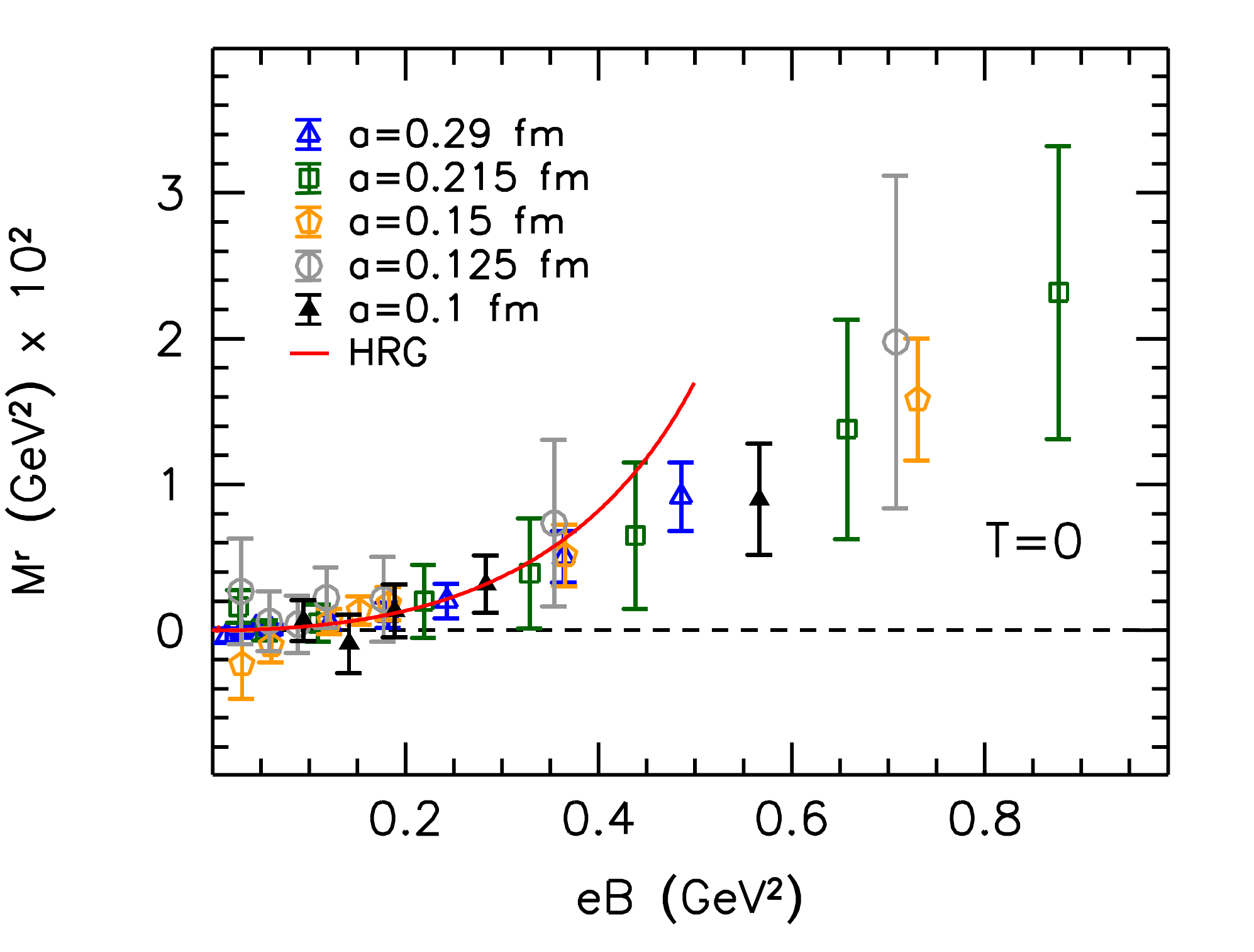}
 \vspace*{-.2cm}
 \caption{ \label{fig:magnetization}
Left panel: up quark contribution to the unrenormalized magnetization at $T=0$, for various lattice 
spacings (colored points). Right panel: renormalized magnetization on the lattice (colored points), and the hadron 
resonance gas (HRG) model prediction (solid red line).
 }
\end{figure}

The contribution of the up quark to $M\cdot eB$ is shown in the left panel of 
Fig.~\ref{fig:magnetization} for several lattice spacings at $T=0$. 
The magnetization is still subject to the additive renormalization as given in Eq.~(\ref{eq:Frenorm}). We observe 
that the renormalized results obtained at five different lattice spacings
agree within their statistical errors, see right panel 
of the same figure. The results indicate that $M^r>0$, i.e. the zero-temperature 
QCD vacuum is paramagnetic. 
The lattice data are also compared here to the hadron resonance gas (HRG) model 
prediction~\cite{Endrodi:2013cs}, revealing 
a nice agreement for $eB\lesssim 0.4 \textmd{ GeV}^2$. 

\section{Magnetization on the lattice -- generalized integral method}
\label{sec:integral}

We proceed by considering an alternative method for the determination of $M^r$, or, equivalently, 
of the dependence of the (longitudinal) pressure $p_z=T/V\cdot \log\Z$ on the magnetic field. 
We consider the integral method, 
where the pressure is written as a multidimensional integral of the partial derivatives of $\log\Z$ 
with respect to the parameters of the theory -- the inverse gauge 
coupling $\beta$, the bare quark masses $m_f$ and the magnetic flux $N_b$. The derivatives with respect to the masses are the quark condensates,
\be
\bar\psi_f\psi_f=\frac{1}{N_s^3N_t}\fracp{\log\Z}{m_f}.
\label{eq:densities}
\ee
Moreover, let us define $\Delta$ as the difference between an observable at $N_b$ and at $N_b=0$. 

\begin{figure}[ht!]
 \centering
 \includegraphics[width=7cm]{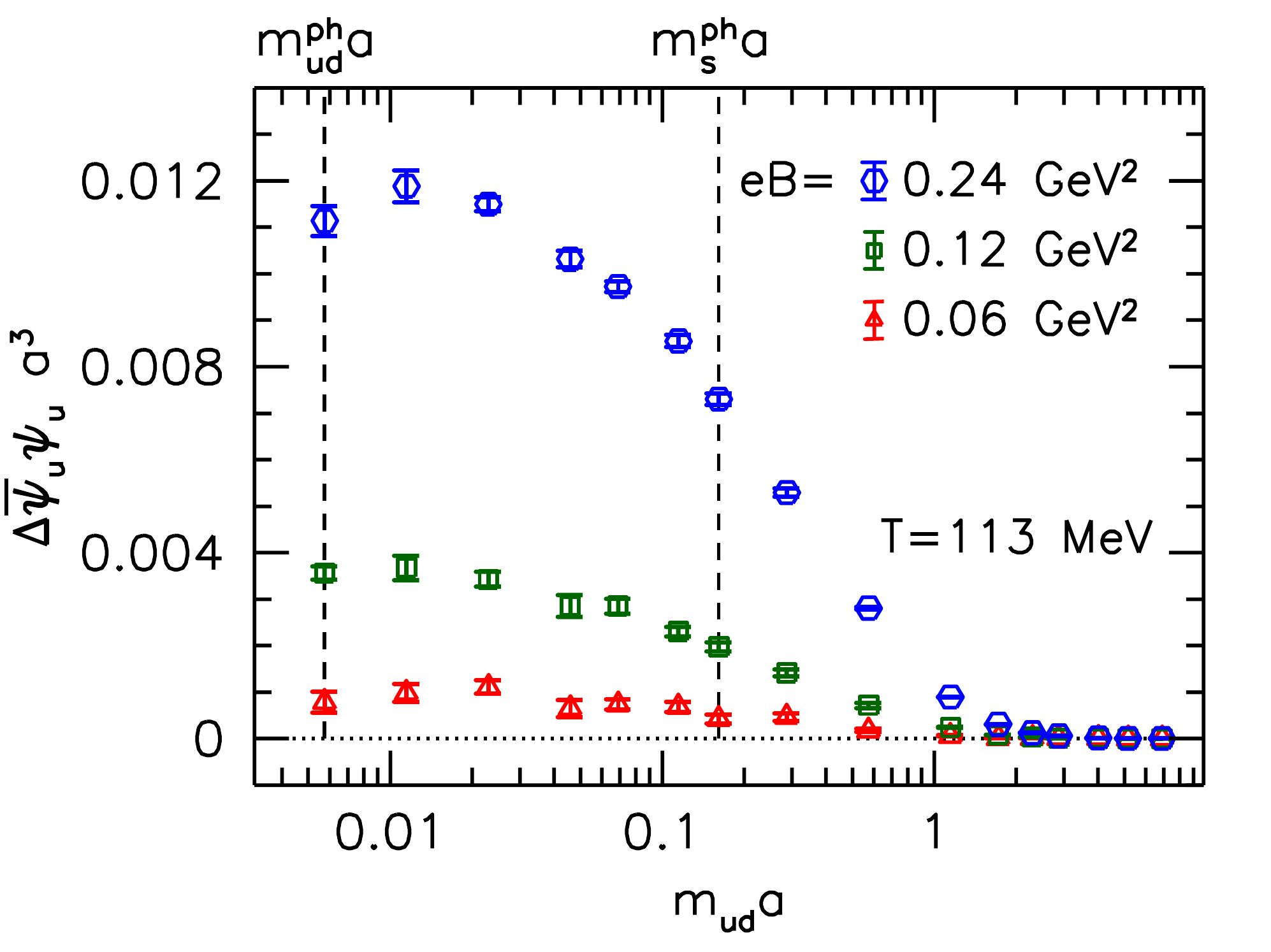} \quad\quad
\includegraphics[width=7cm]{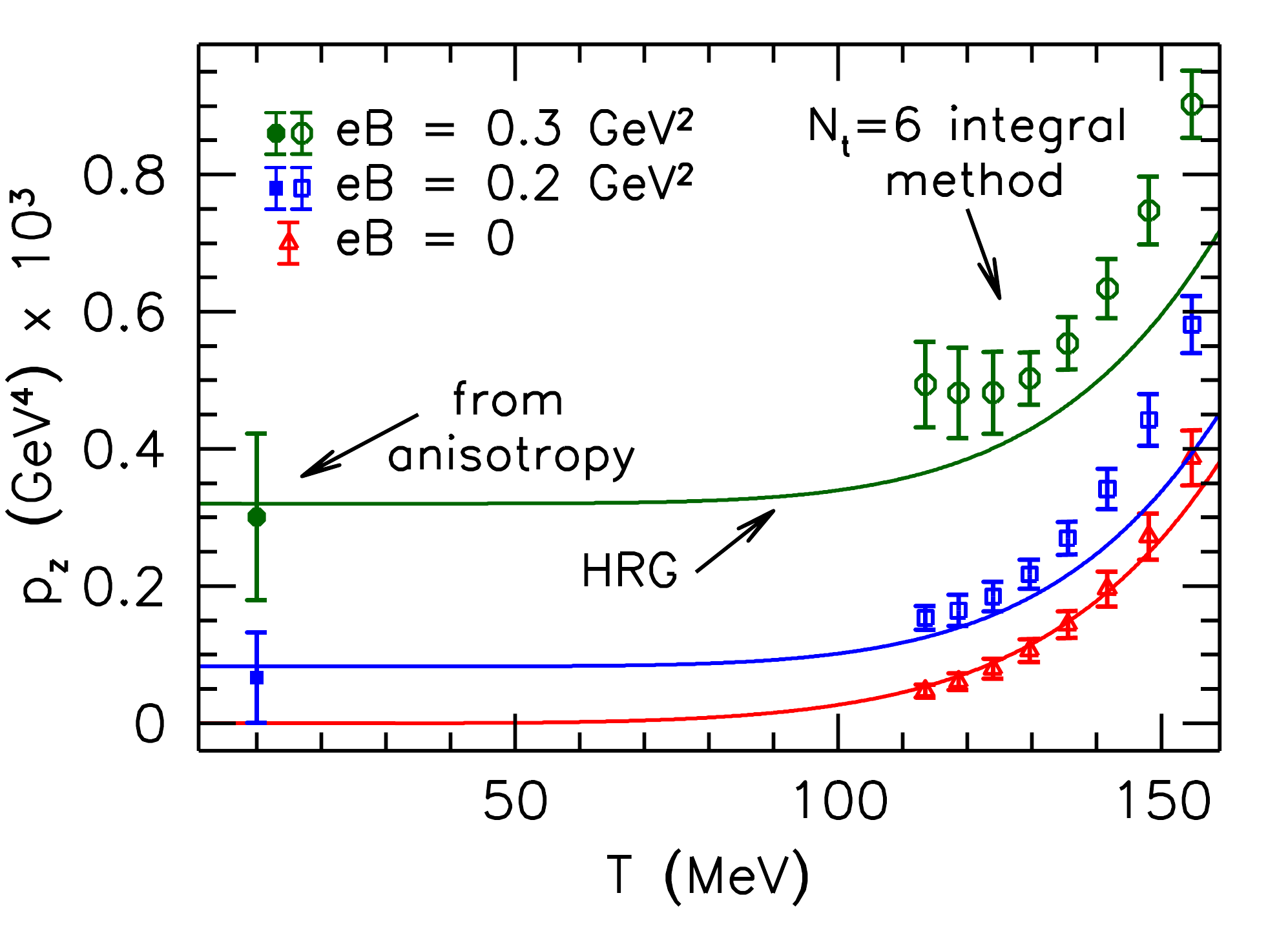}
\vspace*{-.3cm}
 \caption{ \label{fig:deltapbp}
Left panel: change in the up quark condensate induced by $B$ as function of the quark mass
on the $N_t=6$ lattices at $T=113 \textmd{ MeV}$ (different 
colors encode different magnetic fields). Right panel: QCD pressure at 
nonzero magnetic fields from two lattice approaches and the HRG model.
 }
\end{figure}

Flux quantization again represents a conceptual obstacle, as it prohibits 
directly performing the integral on trajectories in parameter space, along which 
the magnetic flux changes. 
To overcome this issue, we propose to consider the trajectory with starting point at
infinitely heavy quark masses. Since QCD with $m_f=\infty$ corresponds to pure 
gauge theory, a magnetic field has no effect here, and thus, $\Delta p_z=0$. Integrating 
down from infinity to the physical quark mass $m_f^{\rm ph}$ allows to extract $\Delta p_z$ at 
any $N_b$ and any $T$ as
\be
\Delta p_z
= -\sum_f \int_{m_f^{\rm ph}}^{\infty} \d m_f \,\Delta \bar\psi_f\psi_f,
\label{eq:pres2}
\ee
which is independent of the integration path. 
In practice we first integrate in the two light sectors up to the symmetric point 
($N_f=3$ theory with different quark charges). Second we integrate in all three sectors up to 
$m_f=\infty$. The contribution of the up quark is shown in the left panel of Fig.~\ref{fig:deltapbp}. 

Once $\Delta p_z$ is known at one $T$ and one $N_b$, 
one can integrate in $\beta$ along a trajectory at that fixed $N_b$ to determine 
the temperature-dependence of $\Delta p_z$.
The renormalization is performed according to Eq.~(\ref{eq:Frenorm}) to 
obtain the renormalized change in the pressure $\Delta p_z^r(N_b)$. An additional 
interpolation is necessary to obtain $\Delta p_z^r(eB)$. 
Shifting this by 
the pressure at $B=0$, which we take from Ref.~\cite{Borsanyi:2010cj}, completes the 
determination of the EoS at nonzero magnetic fields. The so obtained pressure is shown in the 
right panel of 
Fig.~\ref{fig:deltapbp} at low temperatures on our $N_t=6$ lattices, and compared to the anisotropy method 
of Sec.~\ref{sec:aniso} and the HRG model prediction~\cite{Endrodi:2013cs}.
The results consistently show that the pressure is increased by $B$ and, 
accordingly, the thermal QCD vacuum is {\it paramagnetic} 
(moreover, $M^r$ increases as $T$ grows). 
We note that recently two other approaches have also 
been developed to determine the magnetization at $T>0$, giving qualitatively consistent 
results~\cite{Bonati:2013lca,*Levkova:2013qda}. 

\section{Magnetic catalysis, QCD paramagnetism and the QED \boldmath $\beta$-function}

Finally, we would like to point out a relation between three, seemingly 
unrelated phenomena at zero temperature: 1) the magnetic catalysis~\cite{Shovkovy:2012zn} 
of the quark condensates, 
2) the paramagnetism of the QCD vacuum and 3) the positivity of the QED $\beta$-function. 
Let us consider Eq.~(\ref{eq:pres2}) at $T=0$, and apply to both of its sides the 
projection operator $\PP$ 
(see definition in Eq.~(\ref{eq:PPdef})) and $1-\PP$,
\be
\PP[\Delta p_z]
= -\sum_f \int_{m_f^{\rm ph}}^\infty
\d m_f \, \PP [\Delta\bar\psi_f\psi_f] , \quad\;
(1-\PP) [\Delta p_z]
= -\sum_f \int_{m_f^{\rm ph}}^\infty
\d m_f \,
(1-\PP) [\Delta\bar\psi_f\psi_f].
\label{eq:diavspara}
\ee
The left-hand side of the first equation contains the $\mathcal{O}((eB)^2)$ term 
in the zero-temperature pressure, which -- according to our arguments in Eq.~(\ref{eq:Frenorm}) -- is 
just the term that is 
subtracted via charge renormalization and equals $-\beta_1\cdot (eB)^2\cdot \log a^{-1}$ 
(on the lattice $\Lambda=a^{-1}$). 
Would QED be asymptotically free ($\beta_1<0$), the condensate would have to decrease with $eB$
to leading order in some mass regions to make the right hand side positive.
The $\mathcal{O}((eB)^2)$ {\it magnetic catalysis} of the condensate for any $m_f^{\rm ph}$ 
is thus consistent with $\beta_1>0$ 
(which is expected even with 
QCD corrections; the leading perturbative corrections are also 
positive~\cite{Baikov:2012zm}).
Since the lattice regularization provides the upper limit $1/a$ for the integration in the mass, 
this relation also implies that the integral is logarithmically divergent, i.e. 
$\PP [\Delta\bar\psi_f\psi_f]$ decays as $1/m_f$ for large masses.

\begin{wrapfigure}{r}{6.5cm}
 \centering
 \vspace*{-.4cm}
 \includegraphics[width=7cm]{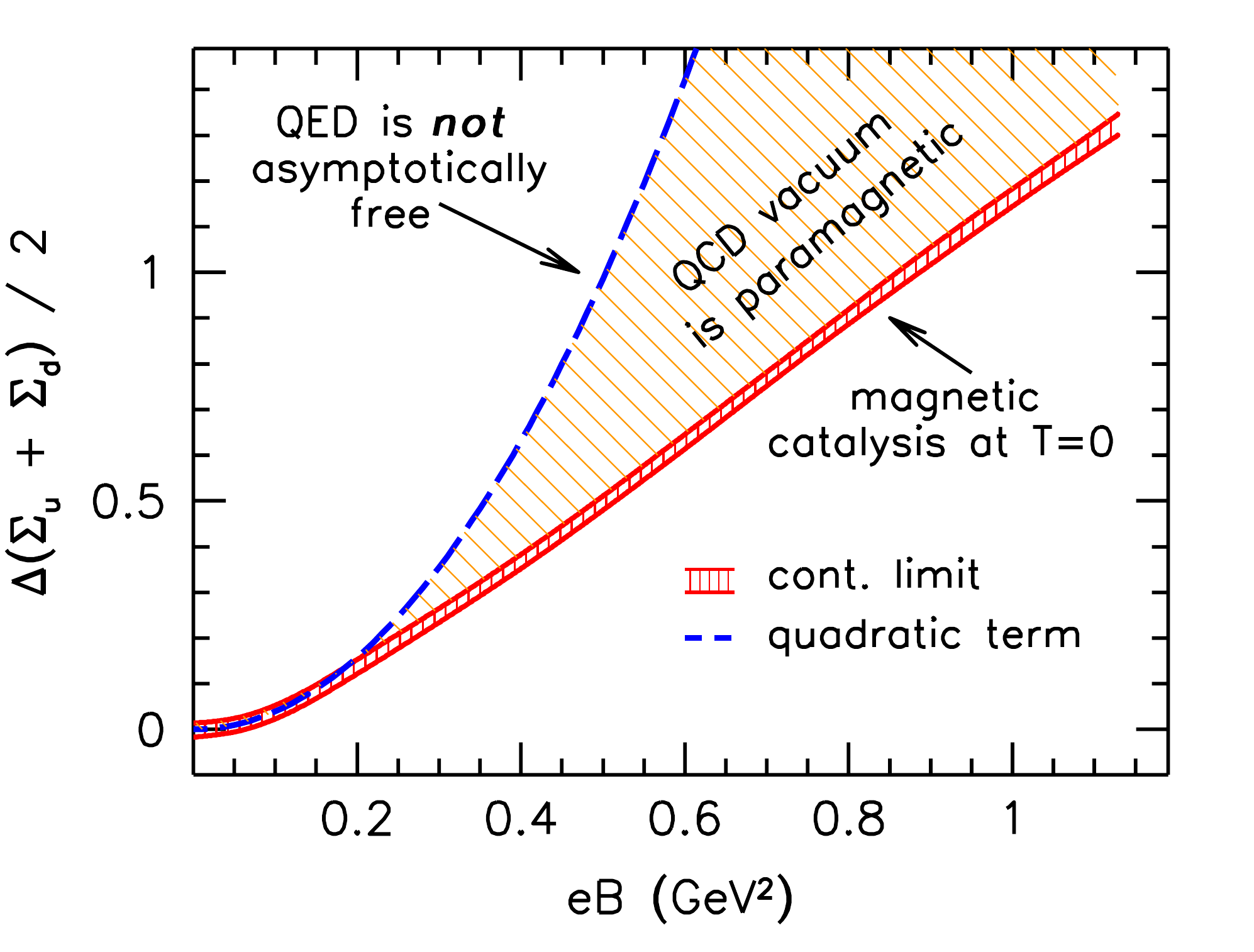}
 \vspace*{-.8cm}
 \caption{ \label{fig:diavspara}
The relation between different concepts: QED $\beta$-function, magnetic catalysis of the condensates 
and QCD paramagnetism.
 }
 \vspace*{-.1cm}
\end{wrapfigure}

Let us now turn to the second relation in Eq.~(\ref{eq:diavspara}). This time, the integrand 
on the right-hand side is the change in the condensate, minus its quadratic part. From our 
lattice measurements of magnetic catalysis we see that the 
condensate as function of $eB$ 
tends to `straighten out' and lies {\it below} its quadratic part~\cite{Bali:2012zg}, see 
Fig.~\ref{fig:diavspara}. 
We also found that this behavior is general for any value of $m_f$. 
Therefore, the right-hand side of the equation is positive. The left-hand side, 
on the other hand, contains the renormalized change in the pressure, 
$\Delta p_z^r$, according to Eq.~(\ref{eq:Frenorm}). 
It also has to be positive, therefore QCD at $T=0$ is 
{\it paramagnetic}. This is indicated by the 
yellow shaded region in Fig.~\ref{fig:diavspara}.

Note that the above arguments concern the behavior of the condensate and of the 
pressure at $T=0$. At finite temperatures both $\Delta p_z$ and $\Delta\bar\psi_f\psi_f$ 
receive finite contributions quadratic in the magnetic field, which are unrelated to charge renormalization. 
In particular, these contributions induce the inverse magnetic catalysis of the 
condensate~\cite{Bali:2012zg,Bruckmann:2013oba} in the transition region $T\approx T_c$, 
which is responsible for the reduction of the chiral crossover temperature with 
growing $eB$~\cite{Bali:2011qj}.

\section{Summary}

In this talk we presented two independent approaches to determine the magnetization 
and the pressure at nonzero magnetic fields. Both the `anisotropy method' and the `generalized 
integral method' indicate the QCD vacuum to be paramagnetic at $T=0$ and at
$T>0$. 
Furthermore, we have shown that the positivity of the QED $\beta$-function is related to 
the $\mathcal{O}((eB)^2)$ magnetic catalysis of $\bar\psi_f\psi_f$ at $T=0$. 
This correspondence was already discussed along similar lines in the HRG model~\cite{Endrodi:2013cs}. 
The negativity of the higher order terms in the condensate, on the other hand, 
establishes the paramagnetic nature of QCD at $T=0$. This is summarized in Fig.~\ref{fig:diavspara}.

\noindent{\\ \bf Acknowledgements. } 
This work was supported by the DFG (SFB/TR 55) and the EU (ITN STRONGnet 238353). G.~E. thanks S\'andor Katz, Igor Shovkovy and K\'alm\'an Szab\'o for enlightening discussions.

\bibliographystyle{JHEP_mcite}
\bibliography{paramag_pos}

\end{document}